\def\year{2021}\relax
\documentclass[letterpaper]{article} 
\usepackage{aaai21}  
\usepackage{times}  
\usepackage{helvet} 
\usepackage{courier}  
\usepackage[hyphens]{url}  
\usepackage{graphicx} 
\usepackage{natbib}  
\usepackage{caption} 
\usepackage{booktabs}
\usepackage{multirow}
\usepackage{multicol}
\usepackage{comment}
\usepackage{xcolor}
\usepackage{dirtytalk}
\usepackage{tabularx}
\usepackage{rotating}
\usepackage{tablefootnote}
\usepackage{hhline}
\usepackage{pbox}
\usepackage[utf8]{inputenc}
\usepackage{amsmath}
\usepackage[T1]{fontenc}
\usepackage{subcaption}

\title{\say{Call me sexist, but...}: Revisiting  Sexism Detection Using Psychological Scales and Adversarial Samples}

\author{
Mattia Samory,
Indira Sen\thanks{Indira Sen and Julian Kohne contributed equally to this work},
Julian Kohne\footnotemark[1], 
Fabian Fl{\"o}ck,
Claudia Wagner\\
}

\affiliations {
Computational Social Science\\
GESIS, Cologne, Germany \\
mattia.samory@gesis.org, indira.sen@gesis.org, julian.kohne@gesis.org, fabian.floeck@gesis.org, claudia.wagner@gesis.org
}

\begin{document}

\maketitle

\begin{abstract}
{Research has focused on automated methods to effectively detect sexism online. 
Although overt sexism seems easy to spot, its subtle forms and manifold expressions are not. In this paper, we outline the different dimensions of sexism by grounding them in their implementation in psychological scales. From the scales, we derive a codebook for sexism in social media, which we use to annotate existing and novel datasets, surfacing their limitations in breadth and validity with respect to the construct of sexism. 
Next, we leverage the annotated datasets to generate adversarial examples, and test the reliability of sexism detection methods. Results indicate that current machine learning models pick up on a very narrow set of linguistic markers of sexism 
and do not generalize well to out-of-domain examples. Yet, including diverse data and adversarial examples at training time results in models that generalize better and that are more robust to artifacts of data collection. 
By providing a scale-based codebook and insights regarding the shortcomings of the state-of-the-art}, we hope to contribute to the development of better and broader models for sexism detection, including reflections on theory-driven approaches to data collection.
\end{abstract}

\section{Introduction}

Sexism is a complex phenomenon broadly defined as ``prejudice, stereotyping, or discrimination, typically against women, on the basis of sex.''\footnote{Oxford English Dictionary} Not only, but distinctively in online interactions, sexist expressions run rampant: one in ten U.S. adults reported being harassed because of their gender,\footnote{\url{https://www.pewresearch.org/internet/2017/07/11/online-harassment-2017/}} and being the target of sexism can have a measurable negative impact \cite{swim2001everyday}. Given the scale, reach, and influence of online platforms,  detecting sexism at scale is crucial to ensure a fair and inclusive online environment.\footnote{https://www.theguardian.com/commentisfree/2015/dec/16/online-sexism-social-media-debate-abuse} Therefore, the research community has been actively developing machine learning approaches to automatically detect sexism in online interactions. Such approaches, on the one hand, provide the foundations for building automated tools to assist humans in content moderation. On the other hand, computational approaches allow at-scale understanding of the properties of sexist language. 

While detecting stark examples of sexism seems relatively straightforward,  \textit{operationalizing and measuring the construct of sexism in its nuances} has proven difficult in the past.\footnote{Constructs are abstract ``elements of information''~\cite{groves2011survey} that a scientist attempts to quantify through definition, translation into measurement, followed by recording signals through an instrument and finally, by analysis of the signals. Constructs may have multiple sub-constructs or dimensions. For example, according to the Ambivalent Sexism theory \cite{glick_ambivalent_1996}, sexism can have two dimensions: benevolent and hostile.} 
Previous work in automated sexism detection focused on specific 
aspects of sexism, such as hate speech towards gender identity or sex \cite{Waseem2016,garibo-i-orts-2019-multilingual}, misogyny \cite{Anzovino2018}, benevolent vs. hostile sexism \cite{Jha2017}, gender-neutral vs. -biased language \cite{Menegatti2017}, gender-based violence \cite{elsherief2017notokay}, directed vs. generalized and explicit vs. implicit abuse \cite{Waseem2017}. The multiple definitions of sexism applied in this related research -- sometimes referring to a sub-dimension of the broader construct -- make it difficult to compare the proposed methods. In particular, a lack of definitional clarity about what aspect a method aims to concretely measure with respect to theory, together with ad-hoc operationalizations, may cause severe measurement errors in the sexism detection models. With few exceptions (e.g. \citeauthor{Jha2017} \citeyear{Jha2017}), previous work tends to neglect or retrofit the link between sexism as a theoretical construct and the data used to train the machine learning models. This impedes assessing which aspects of sexism are contained in the training data. This raises the issue of \textbf{construct validity} \cite{sen2019total,salganik2017bit}. Finally, there have been no detailed investigations into models capturing spurious artifacts of the datasets they were trained on, instead of picking up on more essential syntactic or semantic traits of sexism. This ultimately may give rise to issues of \textbf{reliability} \cite{lazer2015issues,salganik2017bit}, particularly \textit{parallel-forms reliability}, which assesses the degree to which measurements can generalize to variations \cite{knapp2010reliability}.

\textbf{In this work we aim to improve construct validity and reliability in sexism detection.} To achieve this, {our first goal is to map out the various aspects of sexism as a construct as comprehensively as possible for our task of detecting it in natural language. Since no canonical, universally agreed-on list of these aspects exists, we build upon previous work from  social psychology that has implemented a wide variety of operationalizations to measure the construct of sexism (cf. table \ref{tab:codebook}). The survey scales resulting from this work provide us with (i) a workable overview and schema of different categories of sexist attitudes as well as (ii) concrete examples of their materialization in natural language, in the form of survey items.  
The scale items are therefore suitable gold-standard data to built a codebook for sexist content from, as these scales are meticulously designed and tested to ensure construct validity \cite{clark1995constructing}. 
As the dimensionality of the collected scales and covered operationalizations is too high for practical use, we condense them into a codebook with four overarching categories for sexist content (cf. table {\ref{tab:codebook}}) that we extend into a general codebook for sexism by including a additional categories for statements containing sexist phrasing (cf. section \ref{codebook}).} 
The primary purpose of this codebook is to annotate previously-used and novel datasets of online messages in a unified and more nuanced fashion.

Next, following recent advancements in machine learning methods, we leverage the contributions of a large set of crowdworkers to generate adversarial examples{, i.e. examples that are a valid input for a machine learning model, strategically synthesized to put the model to test}~\cite{ilyas2019adversarial}. 
{More recently, the NLP community has also investigated the use of minimal edits to flip the labels of existing data, resulting in `counterfactually-augmented' adversarial examples.}~\cite{kaushik2019learning} 
{In the same vein, }we introduce an approach to generate hard-to-classify examples which have minimal lexical differences from their sexist counterparts, but that are non-sexist. 
This allows us to assess the reliability of existing models.

Our results show that state-of-the-art models fail to generalize beyond their training domain and to broader conceptualizations of sexism. Even in their training domain, the performance of the models severely drops when faced with adversarial examples. Through an analysis of the errors that the machine learning models make, we detail insights on how to build better sexism detection models.

\begin{figure*}
    \centering
    \includegraphics[width=0.8\textwidth]{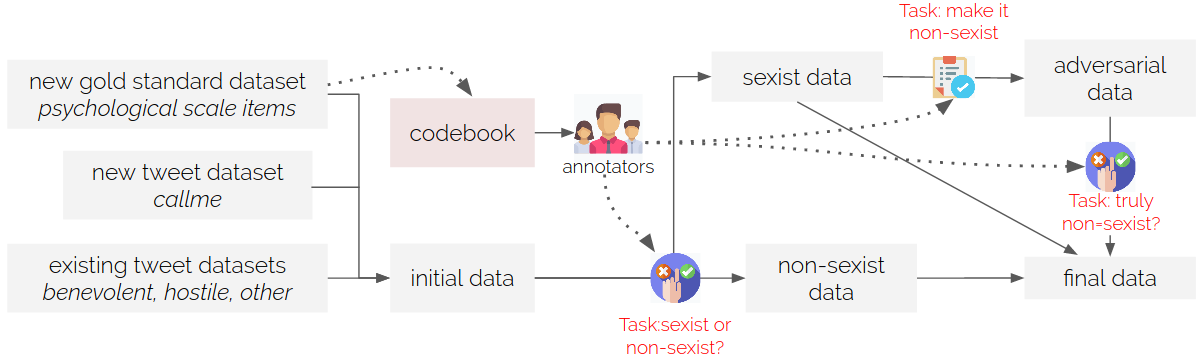}
    \caption{{Data annotation and augmentation pipeline. We start by collecting survey scales used to measure sexism and related constructs, leveraging them (i) to build a codebook for different overarching categories of sexism and (ii) to extract survey items as natural language examples of sexist statements. These items as well as one newly collected  and several pre-existing sets of tweets form our initial data. Our annotators (Mturkers) are trained using the codebook to \textit{identify} sexist and non-sexist examples in the intial data, as well as \textit{generate} adversarial, non-sexist examples by making minimal edits to sexist data points.}}
    \label{fig:datapipe}
\end{figure*}

\section{Data and Methods}

In this section we 
describe how we manually collected and processed items from psychological scales designed to measure sexism  (cf. table \ref{tab:scales}) and developed a sexism codebook  (cf. table \ref{tab:codebook}). We  use the codebook, and an adversarial example generation process, to annotate and expand several datasets  (cf. table \ref{tab:dataset}), {as summarized in figure} \ref{fig:datapipe}.  To conclude, we  detail our benchmark of sexism detection methods.

\subsection{Psychological Scales for  Measuring Sexism}
\label{sec:introducing_scales}
{The first challenge to sexism detection is the lack of definition of sexism that is comprehensive and universally-agreed upon. We tackle this challenge by curating a selection of  psychological scales, designed for measuring sexism and related constructs in individuals.  Our initial selection includes }
scales that were explicitly designed to measure the construct of sexism or are frequently used to measure sexism in the social psychology literature (cf. the top section of table \ref{tab:scales}). We further expanded our selection to include scales that (i) were mentioned in the original selection of sexism scales as related work or (ii) were designed to measure related constructs such as general attitudes towards men or women, egalitarianism, gender and sex role beliefs, stereotypical beliefs about men or women, attitudes towards feminism or gendered norms (cf. table \ref{tab:scales}). 

{The initially selected scales all quantify sexism or a closely related construct. For example, the Attitudes towards Women Scale \cite{spence_short_1973} and the Sex Role Scale \cite{rombough_sexism:_1981} measure attitudes towards the rights and roles of women and men across different domains. In contrast, the Modern Sexism Scale assumes that besides overtly sexist attitudes, sexism can also manifest in more subtle ways like ``denial of continuing discrimination,'' ``antagonism towards women's demands,'' and ``resentment for special favors towards women.'' Similarly, the Neosexism Scale} \cite{tougas1995neosexism} {measures attitudes towards affirmative action policies and perceived overcompensation as proxies for subtle sexist attitudes. Furthermore, the Ambivalent Sexism Inventory \cite{glick_ambivalent_1996} postulates that sexism doesn't have to manifest only in negative attitudes but can also be reflected in benevolent but belittling attitudes, such as women being seen as pure and in need of protection. The Sex-role Egalitarianism Scale \cite{king1997sex} follows a different approach and quantifies egalitarian attitudes instead of sexist attitudes with the premise that these are two different ends of a spectrum. This scale also incorporates items that quantify sexist attitudes against men.\footnote{{While the notion of sexism against men is discussed controversially as it is usually not institutionalized, gender roles and stereotypes also form specific interpersonal expectations towards men. This approach to quantify sexism is taken by many of the existing scales. We stress that this definition is \textit{not} equalling a backlash against feminism and affirmative action is becoming known as ``reverse sexism,'' and we do not endorse this stance in any form.}} Similarly, the Gender-Roles Attitudes Scale \cite{garcia2015development} includes items for measuring stereotypical roles and responsibilities for both men and women as a manifestation of sexist beliefs.}

{While no single scale measures sexist attitudes in their entirety, they all encompass different operationalizations of sexist attitudes with statements that can be considered either sexist or non-sexist. These operationalizations overlap partially, as showcased by the fact that many scales recycle items from previous scales. It thus seems reasonable to gain a better understanding of sexism by incorporating all unique items from the existing sexism scales and categorizing them according to one inclusive schema, providing researchers with a condensed but overarching codebook for basic categories of sexist content (cf. section \ref{codebook}).
}

To be included in our list, the scales had to be publicly available and in English. In addition, items had to be formulated as a subjective, full sentence statement that participants could (dis)agree with. As we were aiming for scale items that are as broad and generalizable as possible with respect to gender, cultural and religious context, we  removed items that refer to specific countries (e.g., ``It is easy to understand the anger of women’s groups in America''), items that were overly specific to a Western cultural or religious context (e.g. ``It is insulting to women to have the ``obey'' clause remain in the marriage service''), items that only apply to one gender or were the gender of the reader influences the perspective on the item (e.g. `` I don't think I should work after my husband is employed full-time''), and items with an overly sexual context (e.g.,``Without an erection, a man is sexually lost''). Due to these criteria, we removed 105 out of 941 items ($\sim$11\%). An additional 72 items ($\sim$8\%) were removed because they were exact or almost exact duplicates of other scale items. In total, the final set thus contains 764 items, of which 536 ($\sim$70\%) measure the construct directly (i.e., a sexist statement) and 228 ($\sim$30\%) via a reverse-coded, non-sexist statement (cf. table \ref{tab:dataset}). 
We confirmed that the selected items are perceived as sexist by annotating each scale item as either ``sexist'' or ``non-sexist'' using five crowdworkers (cf. section \ref{sec:annotation} for details). Overall, the majority verdict of MTurkers corresponded with the ground truth label from the scales for 743 out of 764 items (97.25\%), showcasing that our selection indeed constitutes an accurately labeled dataset of sexist and non-sexist statements.\footnote{The 21 items where the majority verdict did not correspond to the ground truth labels largely comprised items constituting benevolent sexism against women or sexism against men.} Table \ref{tab:scales} gives an overview of all scales we used.

{Our selection of scales covers wider definitional ground than previous work in sexism detection. This allows us to show the limits of current datasets and models. By relying on a broad range of psychological scales that concisely define their respective operationalization of sexism  (cf. table \ref{tab:scales}), this work allows future research to easily identify aspects of sexism still not covered and to expand on them.}

\begin{table*}[!ht]
{\fontsize{9pt}{10.8pt}\selectfont
 \centering  
\begin{tabularx}{\textwidth}{ c|p{9.3cm} X }
\toprule
&\textbf{Scale} & \textbf{Construct} \\
\midrule
{\multirow{7}{*}{\rotatebox[origin=c]{90}{Initial Selection}}}&Attitudes towards Women Scale \cite{spence1972attitudes} & Attitudes towards the role of women in society \\
& Sex Role Scale \cite{rombough_sexism:_1981} & Attitudes towards sex roles \\
& Modern Sexism Scale \cite{swim1995sexism} & Old fashioned and modern sexism \\
& Neosexism Scale \cite{tougas1995neosexism} & Egalitarian values vs. negative feelings towards women \\
& Ambivalent Sexism Inventory \cite{glick_ambivalent_1996} & Hostile \& benevolent sexism \\
& Sex-Role Egalitarianism Scale \cite{king1997sex} & Egalitarian sex-role attitudes (example items) \\
& Gender-Role Attitudes Scale \cite{garcia2015development} & Attitudes towards gender roles \\
 \midrule 
{\multirow{22}{*}{\rotatebox[origin=c]{90}{Additional Scales}}}
 &Belief-pattern Scale for Attitudes towards Feminism \cite{kirkpatrick_construction_1936} & Attitudes towards feminism \\
 &Role conception Inventory \cite{bender_motz_role_1952} & Subjectively conceived husband and wife roles \\
 &Traditional Family Ideology Scale \cite{levinson_traditional_1955} & Ideological orientations towards family structure \\
 &Authoritarian Attitudes Towards Women Scale \cite{nadler_authoritarian_1959} & Authoritiarian attitudes towards women \\
 &Sex-Role Ideology Questionnaire \cite{mason_u._1975} & Women’s sex-role ideology \\
 &Short Attitudes towards Feminism Scale \cite{smith_short_1975} & Acceptance or rejection of central beliefs of feminism \\
 &Sex-role orientation Scale \cite{brogan_measuring_1976} & Normative appropriateness of gendered behavior\\
 &Sex-Role Survey \cite{macdonald_identification_1976} & Support for equality between the sexes \\
 &The Macho Scale \cite{villemez_measure_1977} & Expressions of sexist and egalitarian beliefs \\
 &Sex-Role Ideology Scale \cite{kalin_development_1978} & Traditional vs. feminist sex-role ideology \\
 &Sexist Attitudes towards Women Scale \cite{benson1980development} & Seven components of sexism towards women \\
 &Index of Sex-Role Orientation Scale \cite{dreyer_isro:_1981} & Women’s sex-role orientation \\
 &Traditional-Liberated Content Scale \cite{fiebert_measuring_1983} & Traditional and liberated male attitudes towards men \\
 &Beliefs about Women Scale \cite{belk_beliefs_1986} & Stereotypic beliefs about women \\
 &Attitudes towards Sex-roles Scale \cite{larsen_attitudes_1988} & Attitudes towards egalitarian vs. traditional sex roles \\
 &Male Role Norm Inventory \cite{levant_male_1992} & Norms for the male sex-role \\
 &Attitudes towards Feminism \& Women’s Movement \cite{fassinger_development_1994} & Affective attitudes toward the feminist movement \\
 &Male Role Attitude Scale \cite{pleck_attitudes_1994} & Attitudes towards male gender roles in adolescent men \\
 &Gender Attitudes Inventory \cite{ashmore_construction_1995} & Multiple dimensions of gender attitudes \\
 &Gender-Role Belief Scale \cite{kerr_development_1996} & Self-report measure of gender role ideology \\
 &Stereotypes About Male Sexuality Scale \cite{snell_stereotypes_1998} & Stereotypes about male sexuality \\
 &Ambivalence toward Men Inventory \cite{glick_ambivalence_1999} & Hostile and benevolent stereotypes towards men \\
 \bottomrule
\end{tabularx}}
\caption{Overview of psychological scales measuring sexism and related constructs. The top seven scales represent our initial selection, that we derived our codebook from (cf. section \ref{codebook}). Most scales have multiple subscales that all contribute to operationalizing the respective construct. Items from all scales were selected based on several criteria and the selected ones tested to be perceived as sexist by crowdworkers (cf. section \ref{sec:introducing_scales}). More extensive information about the scales, the items, and our annotations can be found in the supplementary online material.}
\label{tab:scales}
\end{table*}

\begin{table*}
{\renewcommand{\arraystretch}{1.5}
 \fontsize{9pt}{10.8pt}\selectfont 
 \centering  
\begin{center}
\begin{tabularx}{\linewidth}{p{2.5cm} p{5.5cm} X X}  
\toprule
\textbf{Category} & \textbf{Definition} & \textbf{Scale Item Example} & \textbf{Tweet Example} \\ \midrule
 \textit{\textbf{{Behavioral\newline  Expectations}}} & Items formulating a \textit{pre}scriptive set of behaviors or qualities, that women (and men) are supposed to exhibit in order to conform to traditional gender roles &  \textit{
 ``A woman should be careful not to appear smarter than the man she is dating."} & 
 \textit{``Girls shouldn't be allowed to be commentators for football games"}

 \\ 
\textit{\textbf{Stereotypes \&\newline Comparisons}} & Items formulating a \textit{de}scriptive set of properties that supposedly differentiates men and women. Those supposed differences are expressed through explicit comparisons and stereotypes. &
 \textit{``Men are better leaders than women."}
&
\textit{``*yawn* Im sorry but women cannot drive, call me sexist or whatever but it is true."} 

  \\
  \textit{\textbf{Endorsements of\newline Inequality} }& Items acknowledging inequalities between men and women but justifying or endorsing these inequalities. & \textit{``There are many jobs in which men should be given preference over women in being hired or promoted."} & \textit{``I think the whole equality thing is getting out of hand. We are different, thats how were made!"} 
  \\ 

\textit{\textbf{Denying\newline Inequality \&\newline Rejection of\newline Feminism}} & Items stating that there are no inequalities between men and women (any more) and/or that they are opposing feminism & \textit{``Many women seek special favors, such as hiring policies that favor them over men, under the guise of asking for 'equality'."}
    & \textit{``OK. Whew, that's good.  Get a real degree and forget this poison of victimhood known as feminism."} 
\\ 
\bottomrule    
\end{tabularx}
\caption[]{Sexist content categories: we developed the following annotation schema that captures content categories of sexism by manually inspecting items from multiple sexism scales. Note that messages can also be sexist because of phrasing rather than content, as discussed in section \ref{codebook}. All examples of tweets have been editorialized to preserve the privacy of their authors.}
\label{tab:codebook}
\end{center}}
\end{table*}

\subsection{Sexism Codebook}
\label{codebook}

We manually collected, annotated and verified items from psychological scales to (i) define a dataset that allows to assess the validity of automated methods aiming to detect sexism (cf. section \ref{datascales}), and to (ii) derive a codebook for sexism that can be used to annotate social media data.
To develop the codebook, {one of the authors with a background in social psychology} first systematically reviewed the items of our initial selection of survey scales (cf. top part of table \ref{tab:scales}). Then, they determined four major, non-overlapping categories {by iteratively refining common categories} (cf. table \ref{tab:codebook}). Note that these categories are based on multiple scales designed to capture different operationalizations of sexism or closely related constructs at different points in time and tested on different audiences. This enabled us to cover a wide range of aspects of sexism, based on established measurement tools. 

{To test our codebook, we labeled the 91 sexist items from our initial selection of scales (cf. table \ref{tab:scales}) using our four new categories. We asked 5 MTurkers to label them in the same way, with the additional labeling options of ``not sexist'' or ``can't tell.''. We computed a majority verdict for items where at least 3 out of 5 raters chose the same label. This was the case for 78 out of 91 items ($\sim$86\%). Only one item was categorized as ``not sexist,''\footnote{ Item 25 from the Attitudes towards Women Scale: ``The modern girl is entitled to the same freedom from regulation and control that is given to the modern boy.''} and none as ``can't tell.'' The majority verdict coincided with our own annotation for 63 out of 78 items ($\sim$81\%). 
Though, overlap varied across categories, with an almost perfect overlap for the first two ($\sim$92\% and $\sim$100\% respectively), and worse for the latter two ($\sim$43\% and $\sim$53\%). 
Mturkers frequently labeled the items we labeled as an Endorsement of Inequality as Behavioral Expectations, and Denying Inequality \& Rejecting Feminism as Stereotypes \& Comparisons. 
Yet, we had only labeled 7 items as Endorsement of Inequality and 15 as Denying Inequality \& Rejection of Feminism. Consequently, the worse fit only represents few inconsistent items. 
}

{Importantly, one limitation of basing the codebook on scale items is that the scales are designed to measure attitudes rather than behaviors. Therefore, although the categories in table \ref{tab:codebook} help assess statements reflecting underlying sexist attitudes, they may not allow directly assessing other manifestations of sexism specific to online messages. 
In particular, messages can be sexist not only because of \emph{what they say} (e.g. ``In my opinion, women are just inferior to men'') but also because of \emph{how they say it}, even if they do not express a sexist opinion (e.g. ``who asked you? Stupid bitch''). In other words, our scale-based codebook initially only captured statements that are sexist because of their \textbf{content} but not because of their \textbf{phrasing}. 
Therefore, we expand our codebook by adding categories for sexist phrasing, in addition to sexist content. In particular, so as not to conflate sexist phrasing with other forms of incivility (e.g., commonplace profanity), we distinguish sexist phrasing from uncivil statements that are not sexist, and from statements with civil phrasing. The expanded codebook thus also contains the following categories based on phrasing:}

\subsubsection{Uncivil and Sexist Phrasing:} Attacks, foul language, or derogatory depictions directed towards individuals \textit{because of their sex}. The messages humiliate individuals because of their sex or gender identity, for example by means of name-calling (e.g. ``...bitch'') and inflammatory messages (e.g. ``Burn all women!'').

\subsubsection{Uncivil but Non-sexist Phrasing:} Messages that are offensive but not because of the target's sex or gender (e.g.``Are you fucking stupid?'').

\subsubsection{Civil Phrasing:} Neutral phrasing that does not contain offenses or slurs (e.g. ``I love pizza but I hate hotdogs'').

\noindent
{
Note that messages may be sexist just because of their content (e.g. ``Women are inferior to men''), their phrasing (``Basic bitch''), or both (``A slut's place is in the kitchen''). Therefore, we let annotators rate messages along the content and phrasing dimensions independently. For the task of sexism detection we consider a message as sexist if its content and/or phrasing is sexist. Figure \ref{fig:annotation_task} showcases the expanded codebook used for annotating statements as sexist.
}

\subsection{Datasets}
\label{datasets}

Next, we describe the datasets used for annotation, adversarial sample generation, and ultimately sexism detection. 

\subsubsection{Sexism Scales (Abbreviation: \textit{s})}
\label{datascales}
This set contains the scale items described in section \ref{sec:introducing_scales}.

\subsubsection{Hostile Sexism Tweets (Abbreviation: \textit{h})}
\citeauthor{Waseem2016} \shortcite{Waseem2016} used various self-defined keywords to collect tweets that are potentially sexist or racist, filtering the Twitter stream for two months.  The two authors labeled this data with the help of one outside annotator.
In our work we use the portion of the data set labeled as \say{sexist}, and denote it \say{\emph{hostile}}, since according to \citeauthor{Jha2017} \shortcite{Jha2017} it contains examples of hostile sexism. 

\subsubsection{Other (non-sexist) Tweets (Abbreviation: \textit{o})}
\citeauthor{Waseem2016} \shortcite{Waseem2016} also annotated tweets that contained neither sexism nor racism. We denote this dataset as \say{\textit{other}}. 

\subsubsection{Benevolent Sexism Tweets (Abbreviation: \textit{b})}
\citeauthor{Jha2017} \shortcite{Jha2017} extended \citeauthor{Waseem2016}'s dataset to include instances of benevolent sexism: subjectively positive statements that imply that women should receive special treatment and protection from men, and thus further the stereotype of women as less capable \cite{glick_ambivalent_1996}. 
The authors collected data using terms, phrases, and hashtags that are ``generally used when exhibiting benevolent sexism.'' They then asked three external annotators to crossvalidate the tweets to remove annotator bias.

\subsubsection{Call Me Sexist Tweets (Abbreviation: \textit{c})}
To add another social media dataset from a different time period using a different data collection strategy, we gathered data using the Twitter Search API using the phrase ``call me sexist$_(,_)$ but.'' The resulting data spans from 2008 to 2019. For annotation via crowdsourcing, we ran a pilot study and found a statistically significant priming effect of the ``call me sexist$_(,_)$ but'' introduction to tweets on the annotators: if interpreted as a disclaimer, annotators would assume that whatever follows is sexist by default, more so than if the phrase were not there. For this reason we removed the phrase for all annotation tasks producing the output for our eventual analysis, i.e., labeling only the remainder of each tweet (e.g. ``Call me sexist, but please tell me why all women suck at driving.'' to ``please tell me why all women suck at driving'').

\subsection{Annotation and Adversarial Examples}
\label{sec:annotation}
We rely on existing Twitter datasets that were annotated for specific aspects of sexism. To bring them all under a well-defined and consistent scheme we re-annotated them, and the two datasets that we introduce, using our sexism codebook (cf. section \ref{codebook}). 
Then, we challenged crowdworkers to remove as much of the sexism as possible from the sexist examples, so as to produce adversarial examples which have minimal lexical changes, but that are non-sexist. We describe the two crowdsourcing tasks next.

For both tasks, we recruited annotators from Mechanical Turk (\say{MTurkers}). We only accepted annotators located in the US, with over 10,000 HITs approved and over 99\% HIT Approval Rate. Workers also had to pass a strict qualification test that ensured that they understood the construct of sexism as defined by our codebook and didn't apply overly subjective notions of sexism in their labeling. To pass the test they had to correctly annotate 4 out of 5 ground-truth sentences.\footnote{We make available test sentences together with the dataset.} {The mandatory qualification tasks advised MTurkers that the upcoming tasks may contain offensive language (e.g., sexism and swear words). }
\subsubsection{Sexism Annotation Crowdsourcing Task}\label{sec:mturk_annotations}
\begin{figure}
    \centering
    \includegraphics[width=\linewidth]{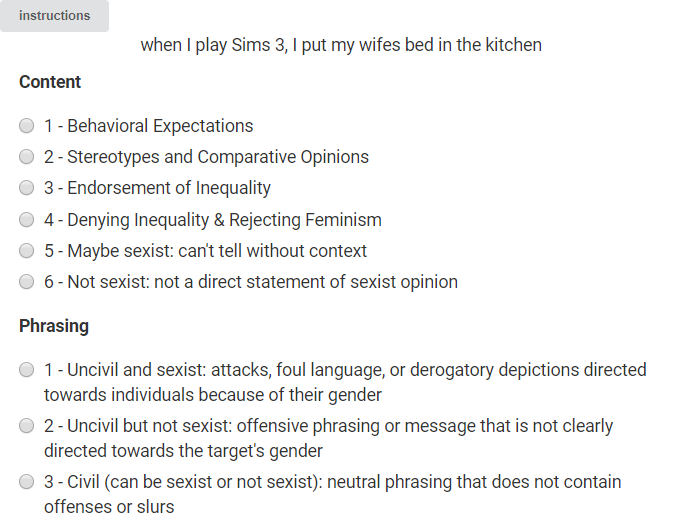}
    \caption{{HIT for annotating sexist statements.}}
    \label{fig:annotation_task}
\end{figure}

We asked MTurkers to annotate if a sentence is sexist (cf. figure \ref{fig:annotation_task}). 
{Specifically, each HIT consisted of one statement that the MTurkers had to annotate on two single-choice lists, containing respectively the codes for sexist content (sexist opinions and beliefs expressed by the speaker) and for sexist phrasing (speaker’s choice of sexist words).}
We referred them to our codebook, together with explanations and examples for each encoding. We paid MTurkers 6 cents per annotation, resulting in a fair hourly wage \cite{Salehi2015WeWorkers}. 
We ran the task on a sample of each dataset (cf. table \ref{tab:dataset}). Five raters annotated each sentence. We marked sentences that at least 3 raters find sexist, because of either content (Randolph $\kappa = .62$) or phrasing ($\kappa = .82$).\footnote{{We had a majority agreement of 81\%, 98.8\% and 100\% for content, phrasing and overall sexism, respectively.}}

\subsubsection{Adversarial Example Generation Crowdsourcing Task} \label{sec:mturk_modifications}
Adversarial examples are inputs to machine learning models that are intentionally designed to cause the model to make a mistake, as they are hardest to distinguish from their counterparts. In the task, MTurkers were presented with sexist messages, and were instructed to \textit{minimally} modify the tweet so that the message is (a) no longer sexist and (b) still coherent.  
For example, the message \say{women are usually not as smart as men} could be amended to ``{women are usually as smart as men}'' or ``women are usually not as tall as men''. 
This task allows us to produce adversarial pairs of texts, which are lexically similar, but with opposite labels for sexism. We paid MTurkers 20 cents per modification. We screened them for task understanding, after showing them examples and guidelines on how to produce meaningful, non-sexist modifications. We asked them not to modify words needlessly (e.g. leave extra non-sexist sentences untouched), unless crucial to make the message coherent. {To better contextualize the message, we also showed whether it had been labeled as sexist because of its content or phrasing, and the majority category. }MTurkers produced one or more modifications for each sexist message (cf. table \ref{tab:dataset}).

As for the original tweets, we also annotated modifications for sexism {(phrasing $\kappa = .96$ and content $\kappa = .86$)}; we discarded all those which were judged sexist {(2.3\%).}\footnote{{Majority agreement for the modifications were 96.5\%, 99.6\% and 100\% for content, phrasing and overall sexism respectively.}}

\begin{table}
{ \fontsize{9pt}{10.8pt}\selectfont \centering  
\begin{tabular}{@{}llllll@{}}
\toprule
\textbf{dataset} & \textbf{\#}& \textbf{w/annot.} & \textbf{sexist} & \textbf{w/mod.} & \textbf{\#mod.} \\ \midrule
benevolent & 678& 678 & 183 & 183 & 396 \\
hostile & 2866& 678 & 278 & 275 & 567 \\
other & 7960& 678 & 8 &  8 & 21 \\
callme & 3826& 1280 & 773 & 765 & 1134 \\
scales & 764& 764 & 536 & 72 & 129 \\ \bottomrule
\end{tabular}
\caption{From each dataset we annotate a sub-sample with our sexism codebook (w/annot.). The sexist samples from the annotated corpora are then modified. The column w/mod. indicates for how many originally sexist samples we were able to obtain modifications, while \#mod. reveals how many modifications were produced in total.} 
\label{tab:dataset}
}
\end{table}
\subsection{Experimental Setup for Sexism Detection}
We use the annotated tweets and their adversarial counterparts for training and testing sexism detection models. We now detail the experimental setup for the task, before introducing the sexism detection models.

\subsubsection{Experimental Setup}
We focus on four dataset combinations. 1) We replicate \cite{Jha2017} by combining the \textit{benevolent}, \textit{hostile}, and \textit{other} datasets (collectively called \textit{bho}). 2) We reproduce the setup using the similar, novel dataset of \textit{callme} tweets  (\textit{c}). 3) We use the \textit{scales} (\textit{s}) dataset as a gold standard test for sexism. 4) To measure the effects of incorporating multiple aspects of sexism, we combine all of the previous datasets into an ``omnibus'' dataset (\textit{bhocs}). We balance classes for training and testing.

To see if modifications aid or hinder sexism detection, we create two types of training data. First we consider the original datasets \textit{bho}, \textit{c}, and \textit{bhocs} as labeled by the MTurkers. Second, we modify the datasets by injecting adversarial examples (while maintaining equal size). In particular, we sample a fraction of the sexist examples, retrieve their modified non-sexist versions, and discard a corresponding number of 
non-sexist examples from the original datasets. We inject adversarial examples so that they make up half of the non-sexist class: this way we ensure that models do not learn artifacts only present in the modifications (such as specific wording used by the MTurkers). We call these modified training sets \textit{bho-M}, \textit{c-M}, and \textit{bhocs-M} respectively.

We construct the \textit{bho}, \textit{c}, and \textit{bhocs} test sets, with the only difference that we substitute the entire 
non-sexist class with adversarial examples in the \textit{bho-M}, \textit{c-M}, and \textit{bhocs-M} test sets. In addition to the tweet datasets, we also test on the \textit{scales} (\textit{s}) dataset. To assess the validity of the sexist detection models, we iteratively sample, shuffle, and split the datasets 5 times in 70\%/30\% train/test proportions.\footnote{We subsample \textit{bhocs} and \textit{bhocs-M} down to 1000 datapoints, to maintain a training set size comparable to that of the other datasets.} 

Next, we introduce the models that we use for sexism detection. We rely on two baselines, (gender word and toxicity) and three models: logit, CNN, and BERT.

\subsubsection{Gender-Word Baseline}
We use the gender-definition word lists from \cite{zhao2018learning} to create a simple baseline. The list contains words that are  associated with gender by definition (like  ``mother'' and ``waitress'').\footnote{The word lists can be found here: \url{https://github.com/uclanlp/gn_glove/tree/master/wordlist}} Our baseline assigns a sentence or tweet to be sexist if it includes at least one gender-definition word. We opt for this method since intuitively, sexist statements have a high probability of containing gender definition-words because of their topical and stereotypical nature. However,
the opposite is not true: the presence of gender-related words does not imply sexism.
\subsubsection{Toxicity Baseline}

We use Jigsaw's Perspective API~\cite{toxicity} -- specifically their toxicity rating -- as another baseline since it is widely used for hate speech detection.\footnote{http://www.nytco.com/the-times-is-partnering-with-jigsaw-to-expand-comment-capabilities/} The Perspective API uses machine learning to measure text toxicity, defined as ``a rude, disrespectful, or unreasonable comment that is likely to make one leave a discussion.'' We determine the optimal threshold differentiating sexist from non-sexist examples during training. We predict those test examples as sexist that score above the threshold.

\subsubsection{Logit}

Our first model is a unigram-based TF-IDF Logistic Regression model with L2 regularization (C = 1). We preprocess text following the procedure detailed in \citeauthor{Jha2017}. While this model is simple, it is interpretable and sheds light on which unigrams contribute to a sentence being classified as sexist.

\subsubsection{CNN}
Our next model consists of a word-level CNN~\cite{kim2014convolutional}, which is a standard baseline for text classification tasks. Each sentence is padded to the maximum sentence length. We use an embedding layer of dimension 128, followed by convolutional filters of sizes 3,4, and 5. Next, we max-pool the result of the convolutional layer into a long feature vector, add dropout regularization, and classify the result using a softmax layer. We adopt the default hyperparameters of 0.5 dropout and no L2 regularization.

\subsubsection{BERT finetuned} 
BERT is a recent transformer-based pre-trained contextualized embedding model extendable to a classification model with an additional output layer~\cite{devlin2018bert}. It achieves state-of-the-art performance in text classification, question answering, and language inference without substantial task-specific modifications. One of the practical advantages of BERT is that its creators make available its parameters after training on a large corpus. BERT can also be adapted for end-to-end text classification tasks through fine-tuning. We replace the output layer of a pre-trained BERT model with a new output layer to adapt it for sexism detection. For preprocessing, all sentences are converted to sequences of 128 tokens, where smaller sentences are padded with spaces. We use default hyperparameters (batch size = 32, learning rate = 2e-5, epochs = 3.0, warmup proportion = 0.1). 

\section{Results}
First, we discuss which aspects of sexism are covered in existing and in our novel social media datasets. 
Next, we compare various sexism detection methods on different datasets, and quantify the usefulness of adversarial examples.

\begin{figure*}
    \centering
    \includegraphics[width=\textwidth]{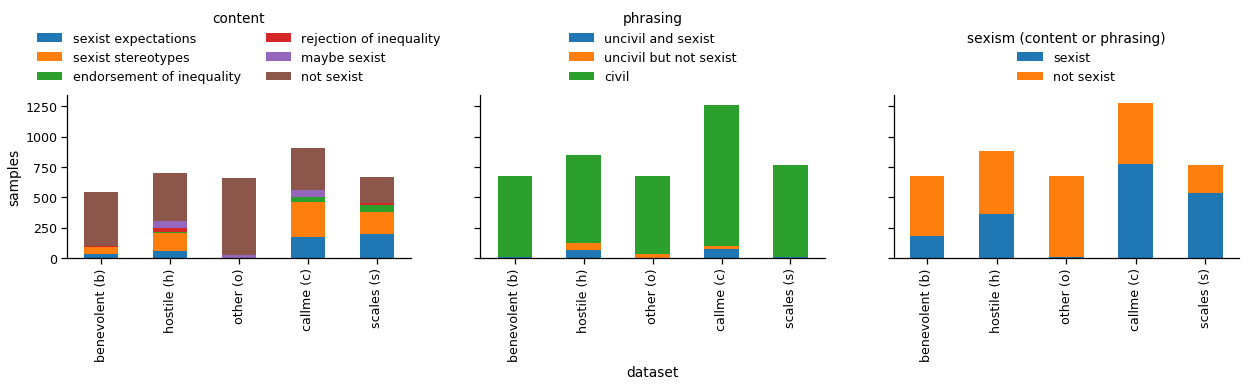}
    \caption{Dataset annotation. We show how the datasets were annotated for different categories of sexist content (left plot) and phrasing (center). We then label messages as sexist if they fall under any sexist content or phrasing category (right plot). We only report messages where at least 3 out of 5 of annotators agree on the label (e.g. for the \textit{callme} dataset, annotators agreed less on content labels than on phrasing labels). 
    The \emph{benevolent} and \emph{hostile} are Twitter datasets that were collected and annotated as sexist in previous work \cite{Jha2017,Waseem2016}. 
    Our re-annotation effort shows that the majority of these tweets were mislabeled and are actually non-sexist. We find that the \emph{other} dataset, that previous work considered non-sexist, does indeed contain largely non-sexist tweets.
    We expected the majority of tweets in the \emph{callme} dataset to be sexist, which proved true ({\textasciitilde}60\%). Finally,  30\%  of the items in our scales dataset are labeled as non-sexist---in line with ground truth.
    }
    \label{fig:annotations}
\end{figure*}

\subsection{Annotating Sexism Datasets}
Figure \ref{fig:annotations} shows which aspects of sexism are covered by different datasets. 
The first subplot shows the number of messages in each dataset for which the annotators agreed upon a category of sexist content. We observe that in all three Twitter datasets (\textit{benevolent}, \textit{hostile}, and \textit{callme}) the largest fraction of sexist tweets exhibits Stereotypes \& Comparisons, followed by Behavioral Expectations. 
{Endorsement of Inequality and Rejection of Inequality are less common on Twitter but still present, especially in the \textit{callme} dataset. Overall, we see that our codebook expands the coverage of various sexism aspects over previous work}~\cite{Waseem2017,Jha2017}. {For example, instances of Rejection of Inequality in the \emph{benevolent} dataset include the tweet, ``Insecure feminist created a day \#adaywithoutwomen  I will not participate because I am a strong, self sufficient woman who doesn't need attn'' which neither falls under gender-based attacks (hostile sexism) nor positive stereotypes (benevolent sexism). The tweet's spurious inclusion in the \textit{benevolent} dataset may be due to the hashtag ``\#adaywithoutwomen,'' signaling positive stereotypes overall, being taken out of context by previous annotators. 
}

The second subplot shows the number of messages for which annotators were able to decide if a message is sexist or not \textit{just based on its phrasing}.  
We find incivility and sexism to be often confused in previous data annotation efforts. For example, focusing on uncivil tweets in the \emph{hostile} dataset, we find that the number of sexist and non-sexist examples is comparable, although the entire dataset was previously labeled as sexist (e.g., ``I really hope Kat gets hit by a bus, than reversed over, than driven over again, than reversed  than..  \#mkr'' or ``I'm sick of you useless ass people in my culture. stfu''). 
Yet, in general we observe that the majority of tweets across all datasets are civil and are therefore either sexist or non-sexist because of their content.

The last subplot combines the prior two and shows the number of sexist and not-sexist messages in each dataset. Remember, a message can be sexist either because of its content or because of its phrasing. 
\emph{Benevolent} and \emph{hostile} are Twitter datasets that were collected and labeled as sexist in previous work \cite{Jha2017,Waseem2016}. Our re-annotation effort with more annotators, who were primed to understand the different aspects of sexism, shows that the majority of these tweets are non-sexist according to our codebook. Interestingly, the \emph{callme} dataset that we compiled with a very simple heuristic, contains the largest fraction of sexist messages out of all Twitter datasets according to our evaluation. The \emph{other} dataset contains tweets that \citeauthor{Waseem2016} labeled as non-sexist. We annotated a random sample of 700 of them, and found that only 8 of them were sexist. Thus, we consider the entire dataset to be non-sexist.

\subsection{Computational Methods for Sexism Detection}

For the sake of brevity, we call ``modified'' the version of the training and test sets including adversarial examples, whereas with ``original'' we denote their counterparts without adversarial examples. We included adversarial examples into the original dataset by randomly substituting 50\% of the non-sexist tweets in the dataset. Therefore, the size of the dataset and the class balance (sexist versus non-sexist) is identical in the original data and the modified data.

\subsubsection{Baseline Performance} We compare models against two baselines: one simplistic, that considers sexist any mention of a gendered word, and one of practical interest, the toxicity score from the Persective API, which is actively used for moderating online content. Our hypotheses on their performance were subverted. Toxicity scores perform the worst across all social media datasets. By analyzing their distribution, we found that the poor performance is due to data topicality. For example, sexist tweets from the benevolent dataset show higher toxicity scores than non-sexist tweets from the same dataset, but lower than non-sexist tweets from the hostile dataset. In other words, toxicity scores may help to correctly classify sexist messages when they are phrased aggressively, but not necessarily when the sexism is expressed in a neutral or positive tone. That is, toxicity might adequately capture sexism that arises through the phrasing of the message, but not because of its content.

Conversely, \textbf{the gender word baseline performed surprisingly well}. 
Somewhat intuitively, when gender is not mentioned, there are fewer chances that a sentence is sexist. 
These results help better understand the potential and limitations of state-of-the-art of sexism detection. Even using the \textit{bho} data and annotations from \citeauthor{Jha2017} (that is, disregarding our own re-annotation efforts), the gender word baseline distinguishes between sexist and non-sexist tweets with 73\% macro F1. This derives from an artifact of data collection and annotation: the two classes are topically distinct and few words easily discern between them---and not because sexism detection is an easy task. Therefore, the performance of a model trained on such data may not reflect its real ability to identify sexist statements in general. This highlights the need for more rigorous model evaluations.

\subsubsection{Overall Performance Patterns} Independently from the type of model employed or from the domain of origin of the training data, several patterns emerge. First, the performance of models trained on original datasets drops drastically when classifying their modified versions (cf. figure \ref{fig:performance_by_train_domain}). For example, Logit classifiers trained on original \textit{bho} data scores around 90\% F1 when classifying original \textit{bho} data, whereas performance drops below 70\% when classifying modified data (faint blue circles, first two rows of the top left plot). This confirms the effectiveness of adversarial examples in challenging the generalizability of the models. In line with our hypotheses, incorporating even only 25\% of adversarial examples into training helps the robustness of the models, which remain the same or improve when classifying modified data (deep blue circles, first two rows of the top left plot). In fact, models trained on modified data outperform models trained on original data even when classifying original data--albeit within statistical error bounds. One reason may be that adversarial examples provide classifiers with more difficult examples and therefore the information gain is higher. Two post-hoc analyses partially support this interpretation. First, by stratifying the misclassification rates, we find that models trained on modified data are more accurate not only on non-sexist adversarial examples, but also on sexist original examples (1--10\% true positive rate increase depending on the dataset). Second, when classifiers are trained on modified data, the most predictive features for the non-sexist class appear qualitatively better aligned with gender-neutral language {(cf. table \ref{tab:feat_importances})}. The original data contain several artifacts that show up as top features in the model such as ``Kat,'' the name of a female TV host: the original data contain personal attacks to her person, which although toxic are often non-sexist. The model trained on modified data instead shows more general features as informative, like ``sexist'' and ``girl'' for positive and ``people'' and ``kid'' for negative predictors.
Thus, \textbf{adversarial examples boost model robustness, and partially, performance}. 

\subsubsection{Best Performance}
We next aggregate performance based on the type of model and data domain used for training. We find that training on the modified version of the omnibus dataset yields the highest average performance across all tests. This suggests that \textbf{incorporating different aspects of sexism in the training set and challenging models with difficult examples helps models generalize better.} Furthermore, the family of models that performs best across tests is BERT. It is the latest and the most complex of the models in this study, therefore it is unsurprising that it achieves peak performance. However, its better average performance may stem from two factors. First, BERT produces document-level representations, thus capturing long-range dependencies in the data. For example, it can summarize a comparison between two genders, regardless of where these appear in a long sentence. Second, BERT's internal embeddings at a word level help deal with lexical sparsity and variety. For examples, this allows associating different expressions of the same stereotype like ``fragile'' and ``delicate''. 

\subsubsection{Performance on Gold Standard Data}
\begin{figure*}[ht!]
    \centering
    \includegraphics[width=\textwidth]{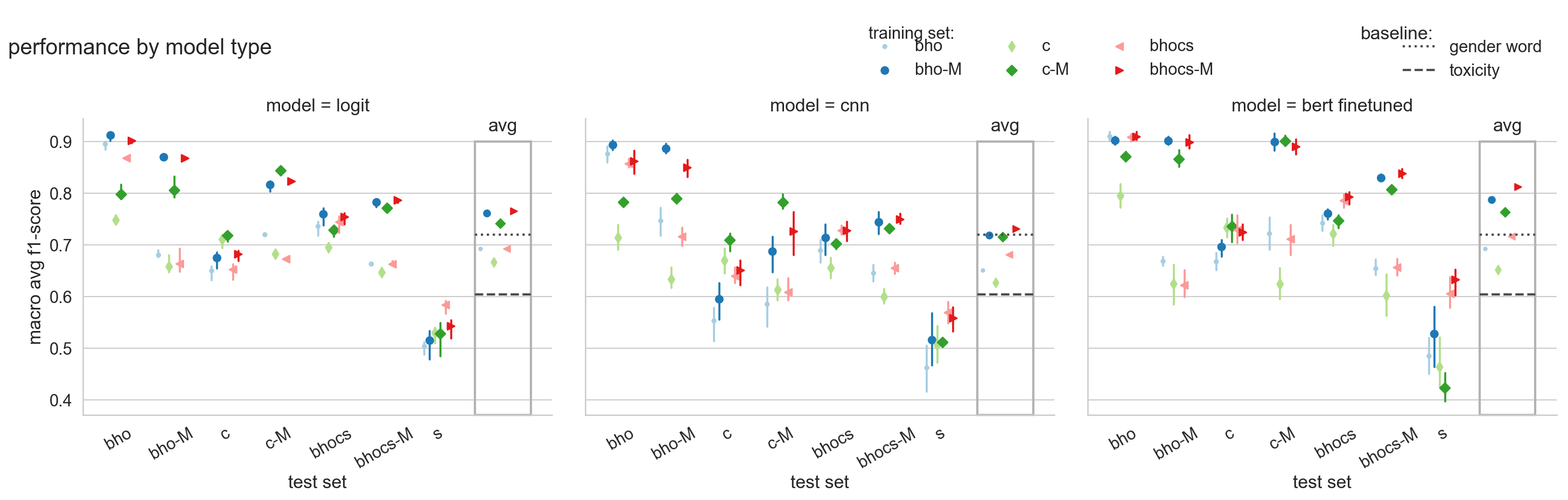}
    \caption{Classification performance of different models. The x axis shows the test data domain and the y-axis depicts the macro-F1 score.
    Deep hues correspond to models trained on modified data, and respectively faint hues to models trained on original data. Colors correspond on the data domain used to train (e.g., blue corresponds to \textit{bho} training sets).
    The rectangle to the right of each plot highlights the average performance across datasets, and compares it to two baselines: the presence of gender words (dotted line) and toxicity score (dashed line).
    Our results show that the fine tuned BERT model that incorporates the adversarial examples into the training process achieves on average the best performance across all datasets. The lowest performance is achieved on the scale dataset (\textit{s}), where the best average F1 score was achieved by the BERT models that were trained on \textit{bhocs-M} ($F1 = 0.63$) and  \textit{bhocs} ($F1 = 0.6$).}
    \label{fig:performance_by_train_domain}
\end{figure*}

Strikingly though, all models perform poorly on the dataset of psychological scales. One potential cause is the different nature of the dataset---survey items, neutral in tone and crafted by academics, are phrased very differently from the social media parlance of the other datasets. However, we find that scales are inherently hard to classify, even when ruling out phrasing issues.{\footnote{{To rule out confounds due to our re-annotation process, we train a logit model on the original labels of the benevolent, hostile and other datasets~\cite{Jha2017}. This model also performs poorly on scales, with an F1 of 42\%, lower than all of the logit models shown in figure~\ref{fig:performance_by_train_domain}}}} The best performance when training classifiers in-domain on the scales themselves, evaluated in cross-validation, is 72\% F1 (compared to over 90\% for \textit{bho}). On the one hand, this speaks for the complexity of the sexism detection tasks. \textbf{Scale items set a hard benchmark to beat for future machine learning approaches}. On the other hand, we find that our best out-of-domain model, BERT trained on the modified omnibus dataset, comes close to in-domain performance at 63\% F1. Intuitively, including data from a variety of sources helps capture the multiple facets of sexism in psychological scale items.

\subsubsection{Error Analysis: When do computational models of sexism fail?}

{
To better understand how different characteristics of a message affect model performance we unpack model errors. We focus on the best performing model, BERT finetuned on the omnibus dataset. We look at both versions of the model: the one trained on the original messages (bhocs), and the one trained on the the adversarial examples (bhocs-M). Then, we assess under which conditions the models misclassify messages. We collect the predictions made for each message in the omnibus test set across validation folds.\footnote{We focus on bhocs-M as it exhibits the most message variety.} Via logistic regression, we predict if the message was correctly classified depending on several factors, controlling for whether the prediction was made by a model trained on original or adversarial data via an interaction term. Do models perform equally well on detecting sexist messages across different content and phrasing categories? To answer that, we include both the binary sexism label as well as the fine-grained labels for sexist content and phrasing obtained through our annotation effort (categorical, dummy coded with ``non-sexist'' as a reference class) in the regression. To assess if message domain plays a role, we also add the dataset of origin of the messages (dummy-coded categorical, ``callme'' as reference). We showed that models trained on adversarial examples are more robust, but it is important that they remain accurate across the various types of messages, and that their performance gain does not stem exclusively from being able to recognize adversarial examples in the test set. Therefore, we also analyze the role of adversarial examples (binary). Finally, expressions of sexism can be subtle, and even human annotators disagree on ambiguous examples. Hence we also include the number of crowdworkers agreeing on the majority label (ordinal). In compact form, the regression is as follows:}
\begin{multline}
correctPrediction \sim adversarialModel *\\ (sexist+content+phrasing+dataset+\\
adversarial+agreement)+\epsilon
\end{multline}
{
We find that model accuracy correlates with the agreement between annotators: messages that are ambiguous for humans are also difficult to classify for computational models. Somewhat intuitively, sexist messages expressing positive stereotypes (i.e. the \textit{benevolent} dataset) as well as on messages that fall under the ``Endorsement of Inequality'' category seem to challenge the models. We also find that training on adversarial examples improves model performance beyond sheer accuracy. Models trained on adversarial data, obviously, better classify adversarial test examples than models trained only on original data. However, the former also better classify harder examples, such as distinguishing between sexist and non-sexist messages with uncivil phrasing. }

\begin{table}[t]

\resizebox{\columnwidth}{!}%
{ \fontsize{9pt}{10.8pt}\selectfont \centering \begin{tabular}{llllllll}

\toprule
\multicolumn{4}{c}{\textbf{original}}&\multicolumn{4}{c}{\textbf{modified}}\\
    \cmidrule(lr){1-4} \cmidrule(lr){5-8}
\multicolumn{2}{c}{\textbf{\textit{positive}}}&\multicolumn{2}{c}{\textbf{\textit{negative}}}&\multicolumn{2}{c}{\textbf{\textit{positive}}}&\multicolumn{2}{c}{\textbf{\textit{
negative}}}\\    
\cmidrule(lr){1-2} \cmidrule(lr){3-4}    \cmidrule(lr){5-6} \cmidrule(lr){7-8}
  \textit{word} &  \textit{$\beta$} &     \textit{word} &   \textit{$\beta$} &    \textit{word} &  \textit{$\beta$} &    \textit{word} &   \textit{$\beta$} \\
    \cmidrule(lr){1-2} \cmidrule(lr){3-4}  \cmidrule(lr){5-6} \cmidrule(lr){7-8}
 sexist &  3.83 &     love &  -1.28 &  sexist &  4.57 &  people &  -1.77 \\
  girl &  2.94 &   people &  -1.11 &    girl &  3.28 &     kid &  -1.26 \\
     rt &  2.89 &      kat &  -1.07 &   women &  3.19 &  person &  -1.12 \\
    man &  2.05 &    happy &  -0.78 &     man &  2.80 &    love &  -1.02 \\
    not &  1.80 &  without &  -0.78 &   girls &  1.63 &  racist &  -0.89 \\
\bottomrule
\end{tabular}}
\caption{{Most predictive unigrams of the sexist (positive) and non-sexist (negative) classes, using the \textit{bho} data and the logistic regression model. The original data (left two columns) contain several artifacts that are picked up by the model such as ``kat,'' the name of a female TV host: the original data contain personal attacks to her person, which although toxic are non-sexist. Vis-\`a-vis, the next two columns show the top unigrams for the corresponding model trained on data containing modifications.}}\label{tab:feat_importances}

\end{table}
\section{Limitations}
{
This work comes with limitations, starting with those stemming from grounding our codebook in psychological scales.
First, even if many of the scales are widely used in contemporary social psychology today, some are dated, raising the concern that new categories of sexism might have developed in recent years and would not be reflected in our selection of scale items. 
However, to the best of our knowledge, more recent validated and well established scales for assessing sexism do not exist. Our work can easily be expanded by future studies, should new scales with additional operationalizations of sexism be published. 

A further note is that sexism scale items are crafted to measure constructs via designed statements. In contrast, social media content does not necessarily reflect the opinion or attitude of the author, evidenced through differences in mode (sarcasm, retweeting, quoting, joking) and context (intent, recipient, previous interactions). By evaluating only the messages themselves, we ignore the wider social context in which they appear. Context has been shown however to alter the subjective perception of whether a statement is sexist or not. These contextual factors include gender of the speaker and recipient \cite{inman1996influence}, 
their relationship} \cite{riemer2014looks,lameiras2018objectifying} {
and the speakers' perceived intent \cite{riemer2014looks}. While a detailed discussion about whether sexism can be evaluated objectively without context or is an inherently context-dependent, subjective phenomenon is beyond the scope of this paper, we argue that there are statements that will be almost universally evaluated as sexist. This argument is supported by the reliability and validity tests during the construction of scales and the strong consensus of MTurkers who labeled 97.25\% of supposedly sexist items from the scales as indeed sexist (cf. section \ref{sec:introducing_scales}). Nevertheless, investigating the influence of contextual factors on the subjective perception of sexism in online messages, as well as of discrimination towards gender identities as more broadly defined, are important avenues for further research. }

{From the point of view of the machine learning task of sexism detection, we use binary classifiers despite our codebook and annotated data would allow us to classify sexism in a more fine-grained manner. While our approach ensures comparability to previous studies, the alternative remains an interesting prospect for future studies. 

Finally, in this work the role of adversarial examples is to assess the shortcoming of current models, and to mitigate the models' lack of reliability by inoculating them in the training sets. However, it is still unclear which data generation strategy would lead to models with tout-court better performance, especially on subtle instances of sexism.
}

\section{Conclusions}

This paper contributes to sexism detection in several ways: First, we proposed a theory-driven data annotation approach that relies on psychological definitions of sexism to uncover the multiple dimensions of the construct. We developed a codebook for sexist expressions on social media, and used it to annotate three existing and one novel dataset/s of tweets. Further, we compiled a gold standard dataset of validated items from psychological scales of sexism.
Moreover, we introduced an approach to generate adversarial examples which have minimal lexical differences from sexist messages. 
Finally, we compared the validity and reliability of state-of-the-art methods for sexism detection. We gave empirical evidence that incorporating adversarial examples as well as multiple data sources boosts model performance across dimensions of sexism. 
Importantly, we showed that two approaches improve model evaluation in sexism detection: 1) challenging model reliability through  adversarial examples;
and 2) confronting the model with the various aspects that comprise sexism as a construct to improve the validity of the model. 
As a benchmark for future approaches to sexism detection, we make available to the research community our data, annotations, and adversarial examples.\footnote{Code available at: \url{https://github.com/gesiscss/theory-driven-sexism-detection} and data at: \url{https://doi.org/10.7802/2251}}

\section*{Acknowledgments}
We are grateful to Anupama Aggarwal for her input on the research design and data collection. 

\fontsize{9.8pt}{10.8pt}\selectfont
\bibliography{main}
\end{document}